\documentclass[preprint,notoc]{JHEP3}
\usepackage{epsfig}
\def\eslt{\not\!\!{E_T}}

\def\to{\rightarrow}

\def\bi{\begin{itemize}}
\def\ei{\end{itemize}}
\def\be{\begin{equation}}
\def\ee{\end{equation}}
\def\bea{\begin{eqnarray}}
\def\eea{\end{eqnarray}}
\def\te{\tilde e}

\def\tu{\tilde u}
\def\tc{\tilde c}

\def\tb{\tilde b}

\def\td{\tilde d}

\def\tst{\tilde t}
\def\ttau{\tilde \tau}

\def\tg{\tilde g}
\def\tnu{\tilde\nu}

\def\tq{\tilde q}
\def\tw{\widetilde W}
\def\tz{\widetilde Z}
\def\alt{\stackrel{<}{\sim}}
\def\agt{\stackrel{>}{\sim}}

\title{
Reconciling Neutralino Relic Density with\\
Yukawa Unified Supersymmetric Models
}

\author{Daniel Auto, Howard Baer, Alexander Belyaev and Tadas Krupovnickas
\\ Department of Physics, Florida State University\\ 
Tallahassee, FL 32306, USA\\
E-mail: \email{auto@hep.fsu.edu}, \email{baer@hep.fsu.edu}, 
\email{belyaev@hep.fsu.edu},\email{tadas@hep.fsu.edu}}
\preprint{\vbox{\hbox{FSU-HEP-040630} }}

\abstract{
Supersymmetric grand unified models based on the gauge group $SO(10)$ 
are especially attractive in light of recent data on neutrino masses.
The simplest $SO(10)$ SUSY GUT models predict unification of third generation
Yukawa couplings in addition to the usual gauge coupling unification.
Recent surveys of Yukawa unified SUSY GUT models predict an
inverted scalar mass hierarchy in the spectrum of sparticle masses if
the superpotential $\mu$ term is positive. In general, such models tend to
predict an overabundance of dark matter in the universe. We survey several
solutions to the dark matter problem in Yukawa unified supersymmetric models.
One solution-- lowering the GUT scale mass value of first and second generation
scalars-- leads to $\tu_R$ and $\tc_R$ squark masses in the $90-120$ GeV
regime, which should be accessible to Fermilab Tevatron experiments. We also 
examine relaxing gaugino mass universality which may solve the
relic density problem by having neutralino annihilations via the $Z$ or $h$
resonances, or by having a wino-like LSP.
}


\begin{document}

\section{Introduction}
\label{sec:intro}

The unification of gauge couplings in the Minimal Supersymmetric
Standard Model (MSSM) can be construed as indirect evidence for
the existence of weak scale supersymmetry. It is of course 
also suggestive of the existence of a grand unified symmetry
at energy scales $Q\ge M_{GUT}\simeq 2\times 10^{16}$ GeV. Such a 
SUSY GUT theory may be the ``low energy effective theory'' that can
be obtained from some more fundamental superstring theory. 
Models based on the gauge group $SU(5)$ are compelling in that they 
explain the apparently ad-hoc
hypercharge assignments of the SM fermions\cite{su5}. 
However, many $SU(5)$ SUSY GUT models
as formulated in four dimensions are already excluded by proton 
decay constraints\cite{pdecay}.
$SU(5)$ SUSY GUT models can also be formulated in five or more dimensions,
where compactification of the extra dimensions leads to a break down 
in gauge symmetry\cite{5dsu5}. These models can dispense with the unwieldy 
large Higgs representations required by four dimensional models, and can also
be constructed to suppress or eliminate proton decay entirely.

The steady accumulation of data on neutrino mass and neutrino oscillations in
the past several years has given strong motivation to the possibility that
nature is described by a SUSY GUT theory based 
on the gauge group $SO(10)$\cite{review,altarelli}.
In $SO(10)$ theories, all 15 of the SM matter fields of a single generation 
inhabit the 16-dimensional spinorial representation of $SO(10)$. The 16th 
element can be taken to be a SM gauge singlet field that functions as
the right-hand neutrino. A large Majorana mass term is allowed for 
this field by the gauge symmetry. Dominantly left handed neutrinos thus gain 
tiny masses via the see-saw mechanism, 
while dominantly right handed neutrinos 
acquire a mass near to the unification scale, and are conveniently swept 
out-of-sight\cite{seesaw}. The neutrino see-saw mechanism also lends 
itself elegantly to a mechanism for baryogenesis via 
leptogenesis\cite{leptogenesis}. For these reasons (and others), 
the possibility that nature is described by an $SO(10)$ SUSY GUT model at
scales $Q\ge M_{GUT}$ deserves serious consideration.

In the simplest $SO(10)$ SUSY GUT models, the two Higgs doublets 
of the MSSM occupy the same
10 dimensional Higgs multiplet $\phi({\bf 10})$. The superpotential 
then contains the term
\be
\hat{f}\ni f \psi({\bf 16})\psi({\bf 16})\phi({\bf 10}) +\cdots
\ee
where $f$ is the single Yukawa coupling for the third generation.
Thus, the simplest $SO(10)$ SUSY GUT models predict 
$t-b-\tau$ Yukawa coupling unification 
in addition to gauge coupling unification. 
It is possible to calculate the $t$, $b$ and $\tau$ Yukawa couplings at
$Q=m_{weak}$, and extrapolate them to $M_{GUT}$ in much the same way
one checks the theory for gauge coupling unification. Yukawa coupling
unification turns out to depend on the entire spectrum of 
sparticle masses and mixings\cite{hrs}
since these enter into the weak scale supersymmetric threshold corrections.
Thus, the requirement of $t-b-\tau$ Yukawa 
coupling unification can be a powerful constraint on the 
soft SUSY breaking terms of the low energy effective theory\cite{old}.

It is well known from early work that large values of $\tan\beta\sim 50$
are required for $t-b-\tau$ unification. However, in models with a common 
scalar mass, such as mSUGRA, it was found that a breakdown in the
mechanism of radiative EWSB occurred which excluded $\tan\beta $ values which
would give rise to Yukawa coupling unification. In Ref. \cite{bdft},
it was found that models with Yukawa coupling unification with $R<1.05$
(good to 5\%) 
could be found if additional $D$-term splittings of scalar masses
were included. 
Here, $R\equiv max(f_t,\ f_b,\ f_\tau )/min (f_t,\ f_b,\ f_\tau )$, where
all Yukawa couplings are evaluated at the GUT scale.
The $D$-term splittings occur naturally when the 
$SO(10)$ gauge symmetry breaks to $SU(5)$, and they are given by\cite{dterms}
\begin{eqnarray}  
m_Q^2=m_E^2=m_U^2=m_{16}^2+M_D^2 , \nonumber \\  
m_D^2=m_L^2=m_{16}^2-3M_D^2 , \nonumber \\  
m_N^2 = m_{16}^2+5M_D^2,\nonumber \\  
m_{H_{u,d}}^2=m_{10}^2\mp 2M_D^2 ,  
\label{eq:so10}
\end{eqnarray}  
where $M_D^2$ parameterizes the magnitude of the $D$-terms.  
Owing to our ignorance of the gauge symmetry breaking mechanism,  
$M_D^2$ can be taken as a free parameter,   
with either positive or negative values.  
$|M_D|$ is expected to be of order the weak scale.  
Thus, the $D$-term ($DT$) model is characterized by the following free   
parameters,  
\begin{eqnarray}  
m_{16},\ m_{10},\ M_D^2,\ m_{1/2},\ A_0,\ \tan\beta ,\ {\rm sign}(\mu ).  
\label{eq:pspace}
\end{eqnarray}  
Using the $DT$ model, Yukawa unification good to 5\% was found
when soft term parameters $m_{16}$ and $m_{10}$ were scanned up to
1.5 TeV, but only for $\mu <0$ values\cite{bdft}. 
The essential quality of the $D$-term 
mass splitting is that it gave the value of $m_{H_u}$ a head start 
over $m_{H_d}$ in running towards negative values, as is required for REWSB.
In Ref. \cite{bbdfmqt}, the $DT$ model for $\mu <0$ 
was explored in more detail, and 
the neutralino relic density was explored. It was found that a good
relic density could be found in the $A$-annihilation funnel\cite{Afunnel},
although a rather large value of $BF(b\to s\gamma )\agt 4\times 10^{-4}$ 
was usually predicted. 
Good relic density was also found in the stau co-annihilation\cite{stau} and 
hyperbolic branch/focus point (HB/FP) region\cite{ccn,fmm,bcpt}, 
but at some cost to the degree of
Yukawa coupling unification\footnote{The HB/FP region is characterized by 
$|\mu |\alt M_2$ so that the LSP is a mixed higgsino-bino, and has a 
large annihilation rate into vector bosons.}.
\TABLE
{
\begin{tabular}{lc}
\hline
parameter    & \hspace*{0.2cm} value (GeV)\hspace*{0.2cm}   \\
\hline
$m_{16}(3)$ & 7826.5 \\
$m_{16}(1)$ & 1202.0 \\
$m_{10}$ & 9652.3 \\
$M_D^2$ & $(2894.9)^2$ \\
$m_{1/2}$ & 80.0 \\
$A_0$ & -16626.0 \\
$\tan\beta $ & 51.165 \\
$f_t(M_{GUT})$ & 0.563 \\
$f_b(M_{GUT})$ & 0.549 \\
$f_\tau(M_{GUT})$ & 0.563 \\
$R$ & 1.02 \\
$\mu$ & 3075.4 \\
$m_{\tg}$ & 380.1 \\
$m_{\tu_L}$ & 1172.2 \\
$m_{\tu_R}$ & 107.0 \\
$m_{\td_L}$ & 1175.0 \\
$m_{\td_R}$ & 1343.7 \\
$m_{\te_L}$ & 693.4 \\
$m_{\te_R}$ & 1734.2 \\
$m_{\tnu_L}$ & 672.7 \\
$m_{\tst_1}$ & 1651.1 \\
$m_{\tb_1}$ & 2330.5 \\
$m_{\ttau_1}$ & 3049.2 \\
$m_{\tw_1}$ &  135.1 \\
$m_{\tz_1}$ & 57.3 \\
$m_{\tz_2}$ & 134.9 \\ 
$m_A$ & 2505.7 \\
$m_h$ & 134.4 \\
$\Omega_{\tz_1}h^2$& 0.093\\
$BF(b\to s\gamma)$ & $3.96\times 10^{-4}  $\\
$\Delta a_\mu    $ & $15.7 \times  10^{-10}$\\
\hline
\label{tab:1}
\end{tabular}
\caption{Masses and parameters in~GeV units for a
Yukawa unified HS model with generational non-universality.
The spectrum is obtained using ISAJET v7.69.
}
}

Later, E821 measurements of the muon anomalous 
magnetic moment appeared, which favored positive values of the superpotential
$\mu$ parameter. In Ref. \cite{bf}, it was found that Yukawa coupling 
unification good to only 30\% could be achieved in $DT$ models with
$\mu >0$ when $m_{16}$ values were scanned up to 2 TeV. The models with the
best Yukawa coupling unification were found to have soft term 
relations
\be
A_0^2\simeq 2 m_{10}^2\simeq 4 m_{16}^2 ,
\ee
which had also been found by Bagger {\it et al.} in the context of
radiatively driven inverted scalar mass hierarchy (IMH) 
models\cite{bfpz,imh,bdqt}\footnote{In IMH models, starting with multi-TeV
scalars masses at $Q=M_{GUT}$, for certain choices of boundary conditions
third generation and Higgs soft masses are driven to sub-TeV values, 
while first and second generation scalars remain heavy.}.
In Ref. \cite{bf}, a model with just GUT scale Higgs mass splittings was
also examined ($HS$), while all other scalars remained universal.
The parameter space of the $HS$ model is that of Eq. \ref{eq:pspace}, 
but where the $D$-term splitting is only applied to the last 
of the relations in Eq.~\ref{eq:so10}.
Yukawa coupling unification in the $HS$ model was found to be comparable 
to the $DT$ model case when $m_{16}$ values up to 2 TeV were scanned.
Finally, in Ref.~\cite{abbbft} (Auto {\it et al.}), 
soft term values of $m_{16}$ up to
20 TeV were explored. In this case, Yukawa unified solutions to better 
than 5\% were found for $\mu >0$ for the $HS$ model when very large 
values of $m_{16}>5-10$~TeV were scanned. The large scalar masses
that gave rise to Yukawa unification also acted to suppress neutralino
annihilation in the early universe, so that rather large values of 
the relic density were found. Models with $\Omega_{\tz_1}h^2<0.2$ could be
found, but only at the expense of accepting Yukawa coupling 
unification to 20\%, rather than 5\%. The models found with low relic density
generally had either a low $\mu$ value, or were in the light Higgs
annihilation corridor, with $2m_{\tz_1}\sim m_h$.

Similar work has been carried on by Blazek, Dermisek and Raby (BDR).
In Ref. \cite{bdr1,bdr2}, the BDR group used a top-down approach to the RG
sparticle mass
solution to find Yukawa unified solutions for $\mu >0$, where they also noted 
that in this case the $HS$ model worked better than the $DT$ model.
In their approach, the third generation fermion masses and other
electroweak observables were an output of the program, so that starting
with models with perfect Yukawa coupling unification, 
they would look for solutions with a low $\chi^2$ value 
constructed from the low energy observables. The BDR Yukawa unified solutions
were also characterized by soft term IMH model boundary conditions.
The solutions differed from those of Ref.~\cite{abbbft} in that 
they always gave a very low value of Higgs mass $m_A$ and also small $\mu$
parameter, indicative of a mixed higgsino-bino LSP. In Ref. \cite{drrr},
the neutralino relic density was examined for the BDR solutions.
Their low $\mu$ and $m_A$ values generally led to very low values of
$\Omega_{\tz_1}h^2$ unless $m_{1/2}$ was small enough compared to 
$\mu $ that the LSP was in the mixed higgsino-bino region.\footnote{
We note that the HS model {\it without} Yukawa coupling unification
has recently been examined by Ellis {\it et al.}, Ref. \cite{ellis_hs}.}

In this paper, we give further scrutiny to the Auto {\it et al.}
Yukawa unified solutions, especially with respect to the relic density of
neutralinos expected from these models. Our goal is to find regions of
parameter space where a high degree of 
Yukawa coupling unification is preserved, but also where a cold dark matter
relic density in accord with WMAP measurements\cite{wmap} 
can be found.\footnote{The WMAP collaboration has determined
$\Omega_{CDM} h^2=0.1126^{+0.0161}_{-0.0181}$ at $2\sigma$ level.} 
To do so, 
we expand the parameter space of the $SO(10)$ model. In Sec.~\ref{sec:2},
we consider relaxing generational universality of the matter scalars. The 
expansion of parameter space in this fashion is consistent with the $SO(10)$
gauge symmetry, which only preserves universality within a generation
(when $SO(10)$ is unbroken). By maintaining TeV scale third generation masses,
but reducing first and second generation masses, light first and second 
generation scalars are produced which enhance the neutralino annihilation rate.
The curious effect in the $HS$ model is that the split Higgs mass contribution
to RG running causes the right up and charm squarks to be 
by far the lightest 
of the scalars, with $m_{\tu_R}$ and $m_{\tc_R}\sim 100$ GeV needed
to achieve the required relic density. We show that such light squarks 
are just at the limit of observability for the Fermilab Tevatron collider.
Our analysis motivates an experimental search for just two species of light
squarks, while all other scalars are in the TeV range, and 
where $m_{\tg}\sim 400$ GeV.

In Sec.~\ref{sec:3}, instead we relax the condition of gaugino mass 
universality. The gauge kinetic function of supergravity theories need 
not adopt its minimal form in order to respect the $SO(10)$ gauge symmetry.
We examine several models found in Ref.~\cite{abbbft}, and vary
the $U(1)$ gaugino mass $M_1$. This allows the bino-like LSP 
to sit atop the $h$ or $Z$ resonance, which in some cases is enough to 
reduce the relic density to the $\Omega_{\tz_1}h^2\sim 0.1$ level.
Also, by increasing substantially the value of $M_1$, ultimately the
lightest neutralino becomes wino-like, and a valid relic density is 
also achieved.
Variation of the gaugino masses preserves the Yukawa 
coupling unification in all the models we examined. We also examine the case 
of lowering the $SU(2)$ gaugino mass $M_2$ to obtain a wino-like LSP.
In this case, the lightest chargino usually drops below limits from LEP2.
In Sec.~\ref{sec:4}, we present a summary and conclusions.
We note here additional studies which have been performed, where
just $b-\tau$ unification was examined\cite{profumo,pallis} 
or where Yukawa quasi-unification was examined\cite{lazarides}.

\section{Scalar mass non-universality and light squarks}
\label{sec:2}

Our goal in this section is to explore whether an expanded parameter
space allowing non-universality of generations can provide new
avenues towards solving the
dark matter problem in Yukawa unified supersymmetric models.
Thus, we expand the parameter space of the $HS$ model to now include
\vskip -0.5cm
\be
m_{16}(1),\ m_{16}(3),\ m_{10},\ M_D^2,\ m_{1/2},\ A_0,\ \tan\beta\ 
{\rm and}\ sign(\mu ) ,
\label{so10space}
\ee
where the matter scalar soft mass is now broken up into
$m_{16}(3)$ for the third generation, and $m_{16}(1)$ for the first
generation. We will also assume $m_{16}(2)=m_{16}(1)$ to reduce
parameter space freedom and to quell potential contributions to
flavor changing processes like $K_L-K_S$ mass difference
and $BF(\mu \to e\gamma )$. Allowing third generation non-universality
can also contribute to the $\Delta m_B\equiv m_{B_H}^0-m_{B_L}^0$
mass difference, the rate $BF(\tau\to e\gamma )$ or $BF(\tau\to \mu \gamma )$,
and $BF(b\to s\gamma )$. However, 
these latter constraints are all rather mild, and in general can allow for 
significant deviations in generational 
non-universality\cite{masiero,misiak,bbkm}.

Our goal is to maintain the successful prediction of Yukawa coupling 
unification found in Ref.~\cite{abbbft} with heavy scalars and an
inverted scalar mass hierarchy, but at the same time reduce some of 
the scalar particle masses so that there is not too severe a suppression
of neutralino annihilation in the early universe which leads to too large a
relic density. The IMH mechanism of Bagger {\it et al.} works if
the mass relation between third generation scalars, Higgses and $A$ terms is
preserved: $A_0^2\simeq 2 m_{10}^2\simeq 4m_{16}^2(3)$. Thus, a possible way
forward is to decrease the value $m_{16}(1)=m_{16}(2)$ so far that a
valid relic density is attained.
It turns out this procedure is possible, but that it leads to
a qualitatively new spectrum of superparticle masses wherein
two of the squarks-- $\tu_R$ and $\tc_R$-- become quite light, 
in the 90-120 GeV regime, while the other scalar masses
remain quite heavy.

A sample spectrum is shown in Table~\ref{tab:1}, which uses Isajet 7.69
to generate the spectrum\cite{isajet}. 
As can be seen, a high degree of Yukawa 
coupling unification is maintained, while simultaneously achieving
an acceptable dark matter relic density $\Omega_{\tz_1}h^2=0.09$.
The low relic density is obtained because the light $\tu_R$ and $\tc_R$
squarks allow $\tz_1\tz_1\to u\bar{u},\ c\bar{c}$ annihilations
to proceed at a high rate due to unsuppressed $t$-channel squark exchange.

Why are some of the squark masses quite light, while most other scalars
are heavy? The answer comes from the 1-loop RGEs for the SUSY soft breaking 
terms. For first generation matter scalars, they have the form\cite{vernon}
\vskip -0.7cm
\bea
\frac{dm_{Q_1}^2}{dt}&=&{2\over 16\pi^2}\left(-{1\over 15}g_1^2M_1^2-
3g_2^2M_2^2-{16\over 3}g_3^2M_3^2+{1\over 10}g_1^2S\right), \\
\frac{dm_{U_1}^2}{dt}&=&{2\over 16\pi^2}\left(-{16\over 15}g_1^2M_1^2-
{16\over 3}g_3^2M_3^2-{2\over 5}g_1^2S\right), \\
\frac{dm_{D_1}^2}{dt}&=&{2\over 16\pi^2}\left(-{4\over 15}g_1^2M_1^2-
{16\over 3}g_3^2M_3^2+{1\over 5}g_1^2S\right), \\
\frac{dm_{L_1}^2}{dt}&=&
{2\over 16\pi^2}\left(-{3\over 5}g_1^2M_1^2-
3g_2^2M_2^2-{3\over 10}g_1^2S\right), \\
\frac{dm_{E_1}^2}{dt}&=&{2\over 16\pi^2}\left(-{12\over 5}g_1^2M_1^2+
{3\over 5}g_1^2S\right),\ {\rm where}\\
S&=& m_{H_u}^2-m_{H_d}^2+Tr\left[{\bf m}_Q^2-{\bf m}_L^2-2{\bf m}_U^2
+{\bf m}_D^2+{\bf m}_E^2\right] .
\eea

In the mSUGRA model, the term involving $S$ doesn't contribute to
RG running at all, since universality implies $S=0$. However, in the
$HS$ model, while the trace over mass matrices remains zero, the contribution
from Higgs splitting is large, especially for our case where the soft Higgs
masses can be $\sim 10$ TeV. Radiative EWSB in Yukawa unified models requires
$m_{H_d}^2 > m_{H_u}^2$, so that in fact $S<0$. Then the soft masses 
with negative co-efficients on the $S$ terms will be highly suppressed in their
RG running, while terms with positive $S$ coefficient will be pushed to 
higher mass values. If the $S$ term dominates the RG running, then indeed 
we expect a large splitting due to RG evolution in all the first and 
second generation soft breaking masses, with the splitting being dictated 
by the soft mass $U(1)_Y$ quantum numbers. The splitting amongst third 
generation soft masses will be much less, since their RGEs include
additional Yukawa coupling times soft mass terms, and the third 
generation soft terms are also large.


In Fig.~\ref{fig:m_5}{\it a}., we show the third generation and Higgs soft 
term running, for the same case as in Table \ref{tab:1}. 
The soft term suppression due to Yukawa coupling 
contributions is apparent. In Fig.~\ref{fig:m_5}{\it b}., we show first 
generation soft term running. In this case, the large splitting due to the 
dominant $S$ term is evident, and causes the $U_1$ mass to run to 
very low values, which gives rise to the light $\tu_R$ and $\tc_R$
squarks. In addition, the $L_1$ soft mass also is suppressed, though not as
much as $U_1$, so that rather light left selectrons and smuons and their
associated sneutrinos are generated.
%
%
%

%

In Fig.~\ref{fig:plane}, we show the $m_{16}(1)\ vs.\ m_{1/2}$ 
plane for fixed $m_{16}(3)=7826.5$ GeV, and other parameters as given in 
Table~\ref{tab:1}. The blue region at low $m_{1/2}$ is excluded
by the LEP2 bound on chargino mass $m_{\tw_1}>103.5$ GeV. The black region 
on the left is excluded because the $m_{U_1}^2$ squared mass 
is driven tachyonic,
resulting in a color symmetry violating ground state. The green region
denotes sparticle mass spectrum solution with relic density
$0.094<\Omega_{\tz_1}h^2< 0.129$, within the WMAP favored regime, while the
yellow region denotes even lower relic density values, wherein
the CDM in the universe might be a mixture of neutralinos plus some other
species. The red region has $\Omega_{\tz_1}h^2<0.5$. It may be allowed
if QCD corrections to $\tz_1\tz_1\to u\bar{u},\ c\bar{c}$ are large and 
positive\footnote{A subset of QCD corrections for
$\tz_1\tz_1\to q\bar{q}$ has been calculated by Drees {\it et al.}, and are 
found to be small\cite{djkn}.}. 
The remaining unshaded regions give sparticle spectra solutions with
$\Omega_{\tz_1}h^2>0.5$, and would be excluded by WMAP. We also show several
contours of $BF(b\to s\gamma )$, $\Delta a_\mu^{SUSY}$ and $R$, the
$max/min$ ratio of GUT scale Yukawa couplings. We see that all constraints
on the sparticle mass spectrum are within allowable limits for the
shaded region to the right of the excluded region.

We have also performed a $\chi^2$ analysis in
the parameter space listed in Eq.~\ref{so10space} and scanned over the
following range of seven parameters:
$$
0<m_{16}(1)<10~TeV,   \ \ 
0<m_{16}(3)<10~TeV,   \ \ 
0<m_{10}<15~TeV,      \ \ 
$$
\begin{equation}
0<m_{1/2}<0.5~TeV,    \ \ 
0<M_D<5~TeV,	       \ \ 
-30<A_0< 30~TeV,       \ \ 
40<\tan\beta<60 .
\label{eq:so10large}
\end{equation}
For every point in parameter space, we calculate the $\chi^2$
quantity formed from $\Delta a_\mu$, $BF(b\to s\gamma )$ and 
$\Omega_{\tz_1}h^2$. 
We present plots with a color coded representation of the $\chi^2$ value, 
where red shading corresponds to high $\chi^2$, green shading corresponds 
to low $\chi^2$ and yellow shading has intermediate values. 
The numerical correspondence is listed in Fig. \ref{fig:so10_scan}.
As a result, we have found that in order to have Yukawa unification below $\sim 5$\% level
and low $\chi^2$, one should be in the special,
strongly correlated region of the parameter space. For further analysis we 
have fixed some parameters at the values favored by the $\chi^2$ analysis:
\begin{equation}
m_{16}(3)=7750~\rm GeV\it,\quad A\rm_0=-2.05~\it m\rm_{16}(3),
\end{equation}
and varied the others within the $\chi^2$ fit preferred range:
\begin{eqnarray}
49.8\le\tan\beta\le 51.6,\quad 75~\rm GeV\le\it m\rm_{1/2}\le 120~GeV,\nonumber \\
\quad 0.14\le\it\frac{m_{H_d}-m_{H_u}}{m\rm_{16}(3)}\le\rm 0.23,\quad
1.225\le\it\frac{m_{H_d}+m_{H_u}}{\rm 2\it~m\rm_{16}(3)}\le\rm 1.27.
\label{eq:so10fine}
\end{eqnarray}
Finally, for each point generated in the range specified by 
Eq.~\ref{eq:so10fine} the value of $\chi^2$ was minimized by varying 
$m_{16}(1)$.

%

Those solutions and respective correlations
among the model parameters are illustrated in Fig.~\ref{fig:so10_scan}
where we show the results of scanning over the complete
parameter space listed in Eq.~\ref{eq:so10large}. 
The color coding of the points 
corresponds to a $\chi^2$ value computed from the value of $\Omega_{\tz_1}h^2$,
$\Delta a_\mu^{SUSY}$ and $BF(b\to s\gamma )$. In constructing the $\chi^2$
quantity, we adopt the limits that $\Omega_{\tz_1}h^2 =0.1126\pm 0.00805$
(we use the upper limit only), $\Delta a_\mu^{SUSY}= (27\pm 10)\times 10^{-10}$
(as required by the recent E821 $(g-2)_\mu$ measurement compared 
to SM predictions\cite{gm2}), and 
$BF(b\to s\gamma )=(3.25\pm 0.54)\times 10^{-4}$\cite{constraints}.

In Fig.~\ref{fig:dots1}, we show the results of scanning over the
narrowed parameter space defined by Eq.~\ref{eq:so10fine}
in the
$R\ vs.\ \Omega_{\tz_1}h^2$ plane with the same color coding as in Fig.~\ref{fig:so10_scan}.
We see that a large range of points 
are generated with low $\chi^2$ value, and that many of these have 
Yukawa coupling unification below the $5\%$ level, and also with 
a WMAP allowed relic density.

\FIGURE{\epsfig{file=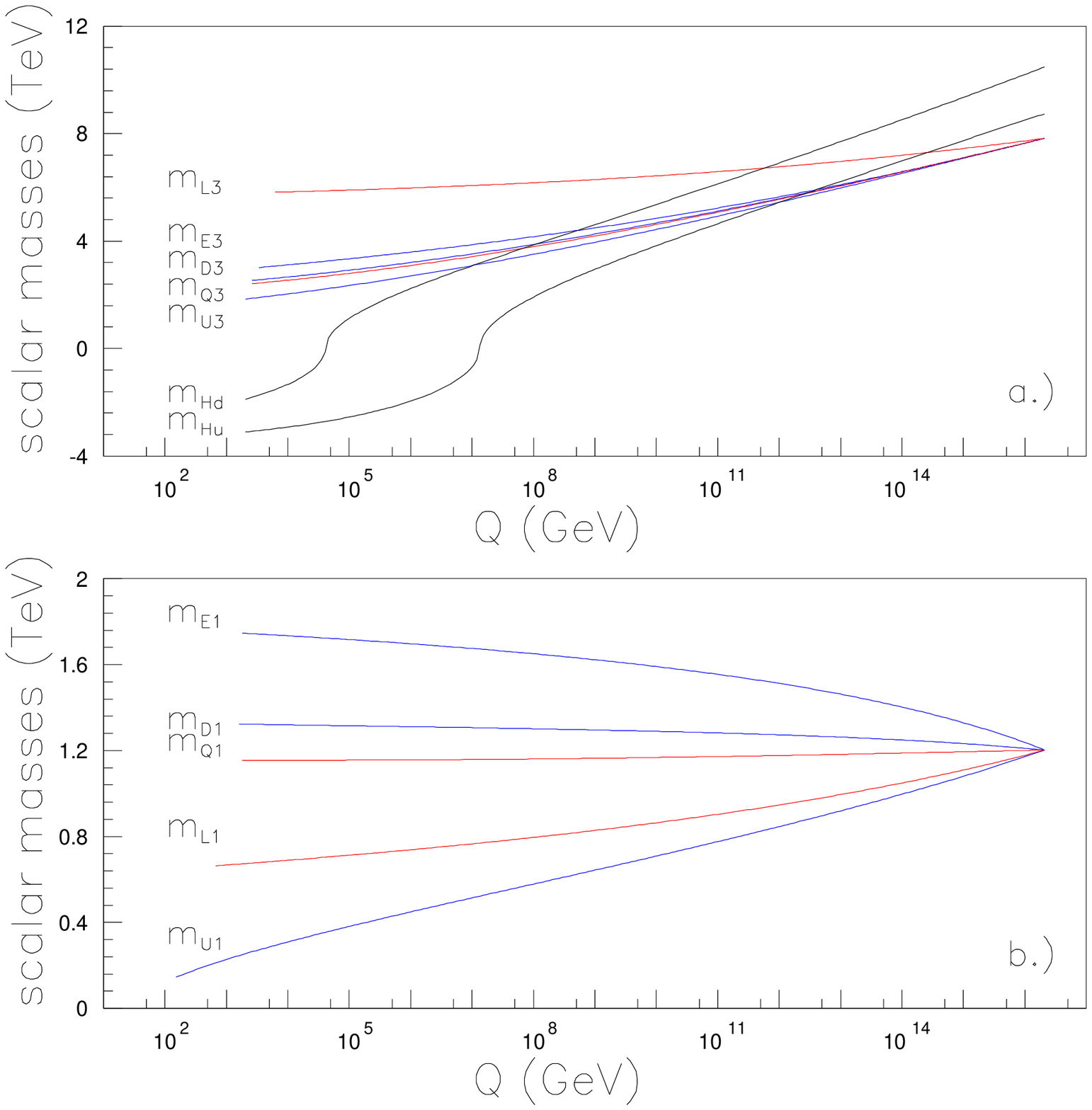,width=14cm} 
\vspace*{-0.8cm}
\caption{Evolution of {\it a}.) third generation and Higgs scalar
soft masses and {\it b}.) first generation soft masses for the case of
$m_{16}(3)=7826.5$ GeV, 
$m_{10}=9652.3$ GeV, $M_D=2894.9$ GeV, $m_{16}(1)=1202.3$ GeV, 
$m_{1/2}=80$ GeV,
$A_0=-16626$ GeV, $\mu >0$, $\tan\beta =51.165$
and $m_t=180$ GeV. For Higgs soft masses, we actually plot 
$sign(m_H^2)\cdot\sqrt{|m_H^2|}$.
}
\label{fig:m_5}
}

\FIGURE{\epsfig{file=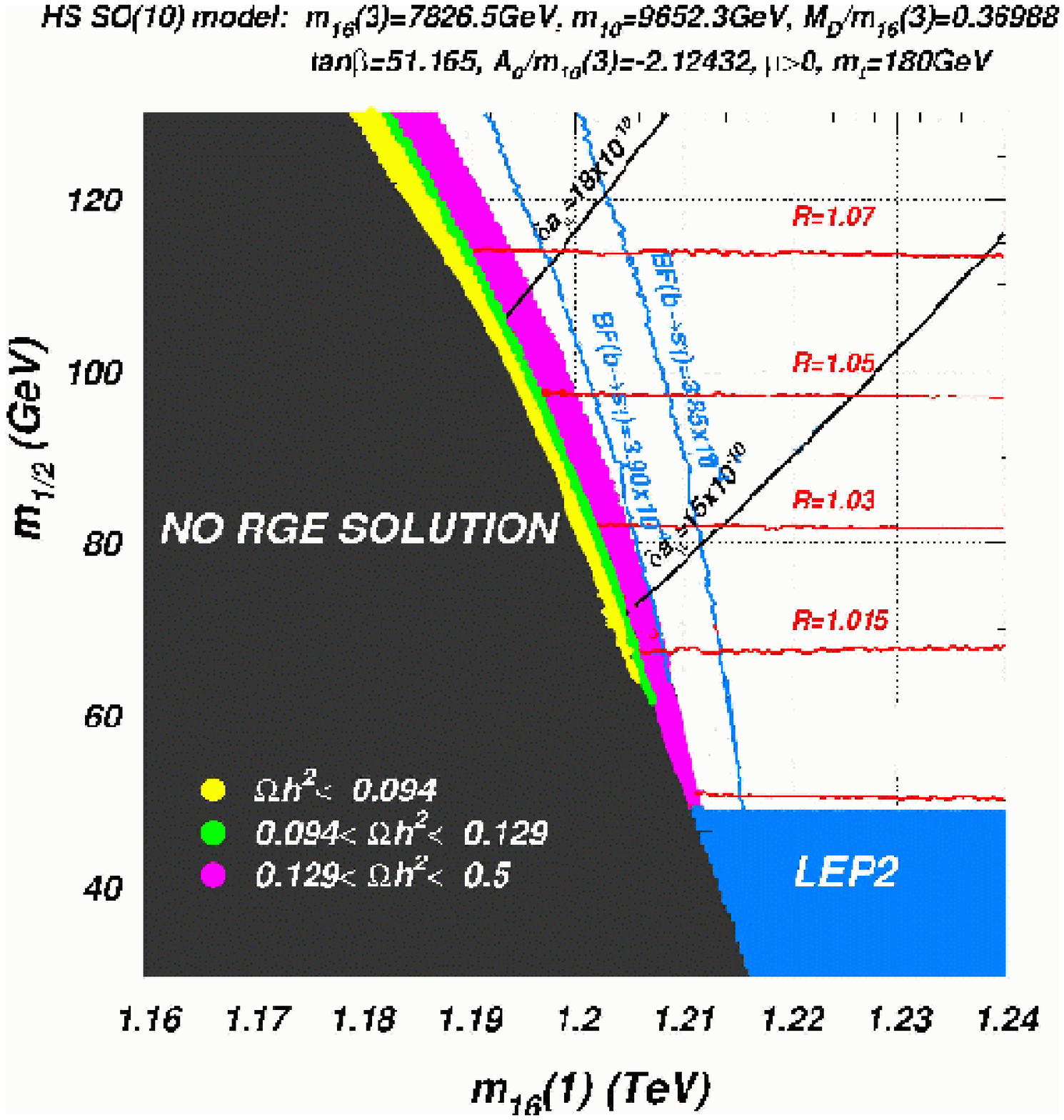,width=14cm} 
\vspace*{-0.8cm}
\caption{Allowed parameter space of Yukawa unified  supersymmetric 
HS model with generational non-universality. We show the
$m_{16}(1)\ vs.\ m_{1/2}$ plane for $m_{16}(3)=7830$ GeV, 
$m_{10}=9650$ GeV, $M_D/m_{16}(3)=0.37$, $A_0/m_{16}(3)=-2.1$, 
$\mu >0$, $\tan\beta =51$
and $m_t=180$ GeV.
The black shaded region gives tachyonic particles, while the blue region is 
excluded by LEP2 chargino search experiments. The yellow and green
regions are allowed by the WMAP determination of $\Omega_{\tz_1}h^2$. We also
show contours of $R$, the measure of Yukawa unification at $M_{GUT}$.
}
\label{fig:plane}
}

\FIGURE[htb]{
\epsfig{file=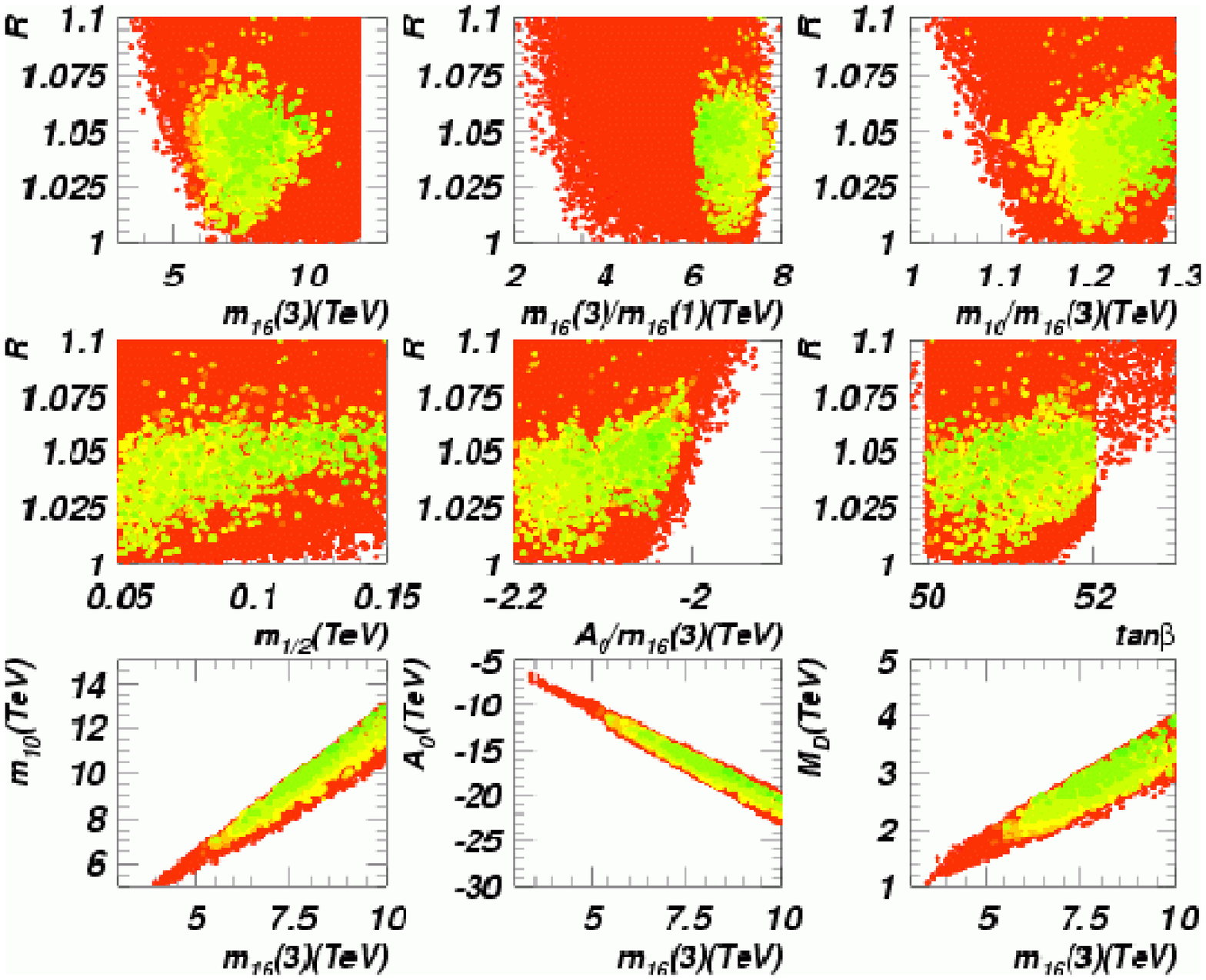,width=14cm}\hspace*{5cm}\\
\epsfig{file=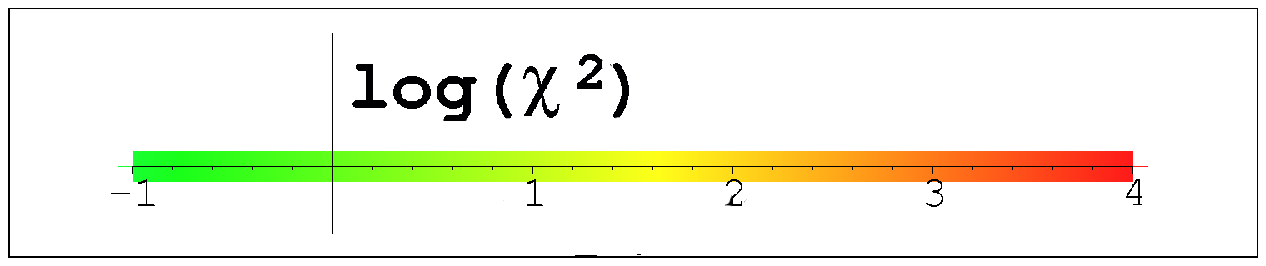,width=14cm} \hspace*{5cm}\\
\caption{
Scan of $SO(10)$ $HS$ models with generational non-universality
for various planes of the parameter space.
The colors of the dots
correspond to the $\chi^2$ value computed from
$\Omega_{\tz_1}h^2$, $BF(b\to s\gamma )$ and $a_\mu^{SUSY}$. 
}
\label{fig:so10_scan}
}

\FIGURE{\epsfig{file=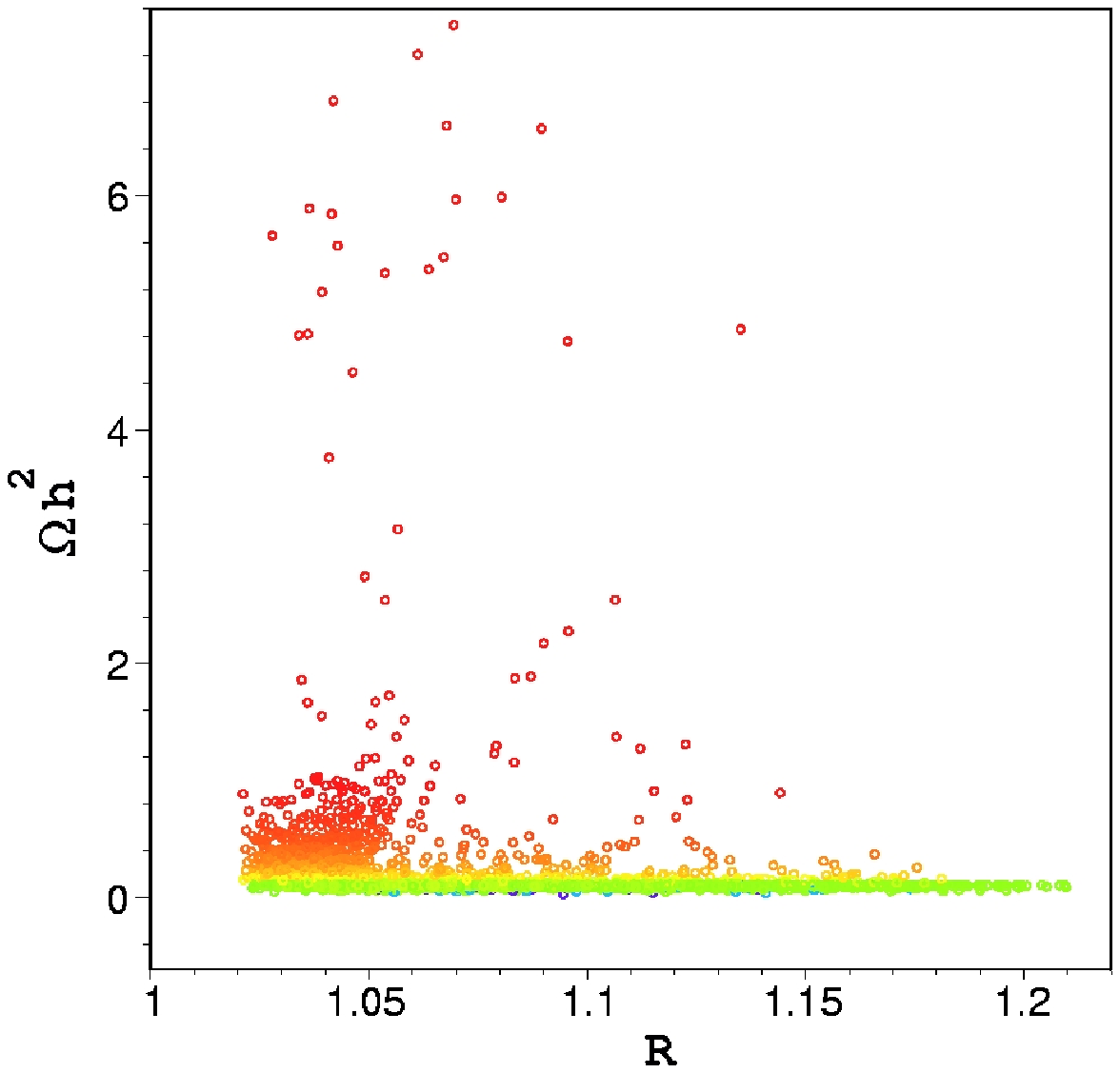,width=10cm}\\
\epsfig{file=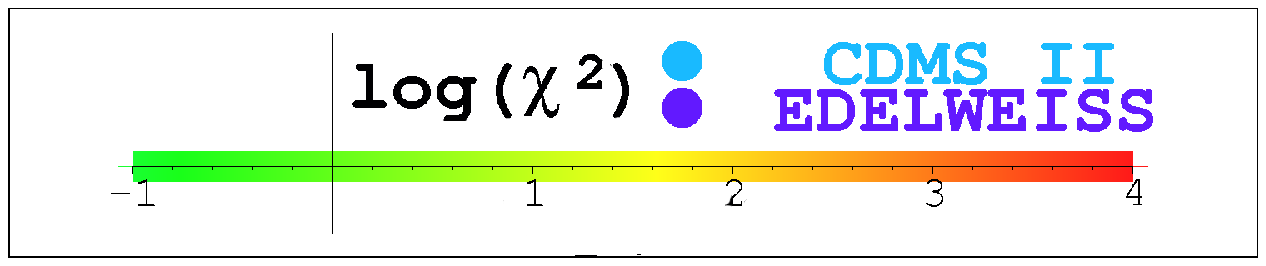,width=10cm} 
\vspace*{0.5cm}
\caption{Scan of $SO(10)$ $HS$ models with generational non-universality,
in the $R\ vs.\ \Omega_{\tz_1}h^2$ plane. The colors of the dots
correspond to the $\chi^2$ value computed from
$\Omega_{\tz_1}h^2$, $BF(b\to s\gamma )$ and $a_\mu^{SUSY}$. 
Points excluded by Edelweiss (CDMS II) direct dark matter searches are
denoted by purple (blue) dots.
}
\label{fig:dots1}
}

In Fig.~\ref{fig:dots2}, we show the same points from the parameter space
scan, but this time in the $m_{\tu_R}\ vs.\ R$ plane. In this case, we
see that essentially all of the low $\chi^2$ points have $m_{\tu_R}$
values in the 90-120 GeV range. For models with $m_{\tu_R}>120$ GeV, 
the value of $\Omega_{\tz_1}h^2$ exceeds the WMAP bounds, and the models
become excluded. The allowed models shown all have $R$ values in the
1.02-1.21 range. 
\FIGURE{\epsfig{file=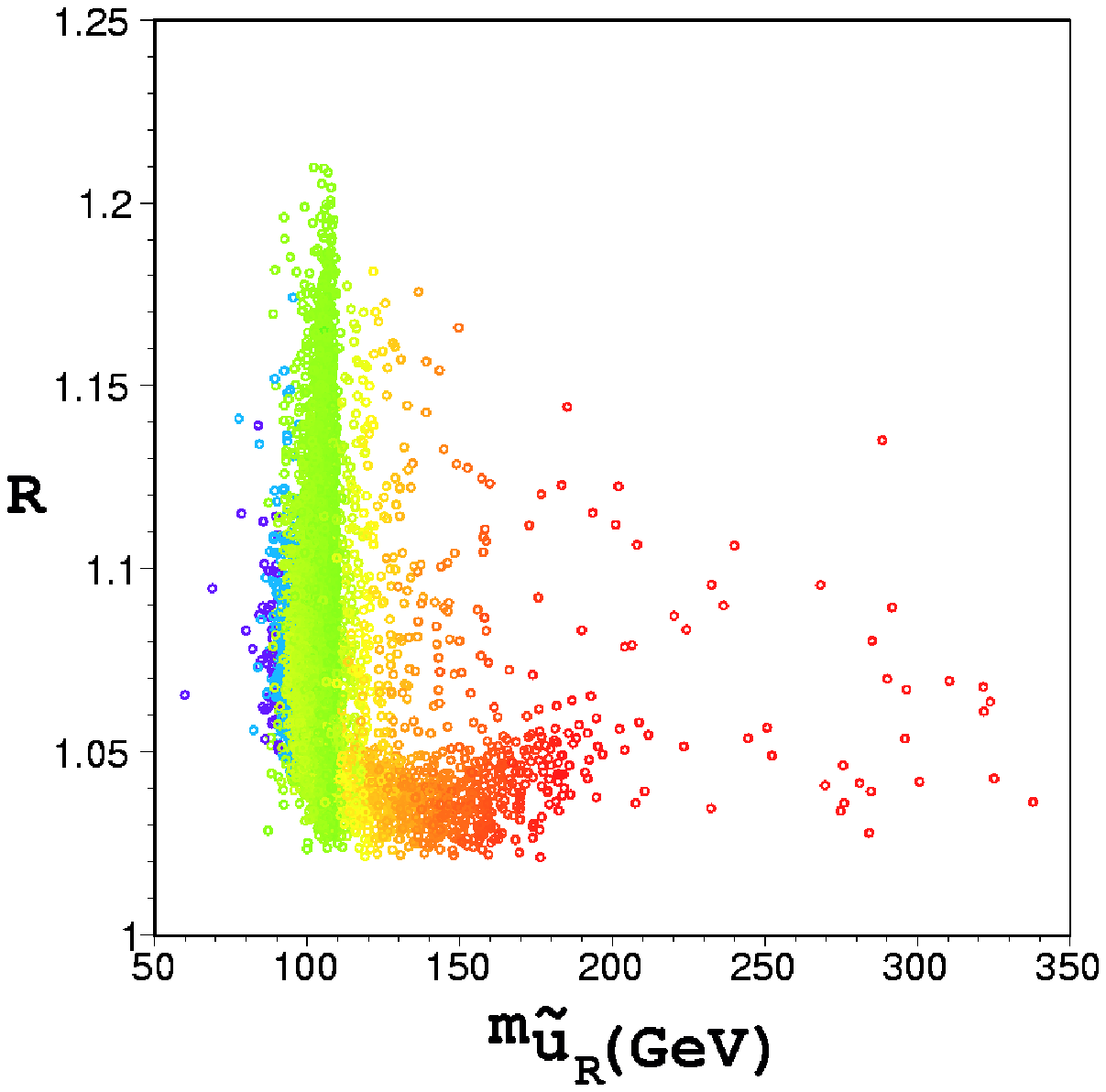,width=10cm}
\vspace*{0.5cm}
\caption{Scan of $SO(10)$ $HS$ models with generational non-universality,
in the $R\ vs.\ m_{\tu_R}$ plane. The colors of the dots
correspond to the $\chi^2$ value as listed in Fig.~\ref{fig:dots1}.
}
\label{fig:dots2}
}

\subsection{Phenomenological consequences of two light squarks}

Given that the Yukawa unified model with split generations predicts
two light squarks of mass 90-120 GeV in order to obtain a relic density 
in accord with WMAP, it is natural to examine the phenomenological 
consequences of such light squarks. We first turn to data from LEP2.
In Ref. \cite{aleph}, the ALEPH collaboration reports mass limits
on squarks in the MSSM. Using 410 pb$^{-1}$ of data ranging up to
CM energies of 201.6 GeV, they report $m_{\tq}>97$ GeV at 95\% CL
assuming $\tq\to q\tz_1$ with a mass gap $m_{\tq}-m_{\tz_1}>6$ GeV, 
and assuming 10 species of degenerate squarks. In our case, with just two 
nearly degenerate squarks, the mass limits will be even 
lower\footnote{If we make the simplifying assumption that all of the 
mass limit comes from running at the highest energy of $\sqrt{s}=201.6$ GeV, 
then the 97 GeV mass limit for 10 degenerate squarks would be reduced
to 89 GeV if only $\tu_R\bar{\tu}_R$ and $\tc_R\bar{\tc}_R$ are
produced.}.
Furthermore, the OPAL collaboration\cite{opal} 
has searched for $\tst_1\bar{\tst}_1$
pairs assuming $\tst_1\to c\tz_1$. These limits exclude can be applied
to our case of $\tc_R\bar{\tc}_R$ production, and exclude
$m_{\tc_R}<90$ GeV for almost all values of $m_{\tz_1}< m_{\tc_R}-m_c$.

There are also limits on squark pair production from Fermilab Tevatron 
experiments. In Ref. \cite{cdf}, the CDF collaboration reports a 
mass limit of $m_{\tq}\agt 250$ GeV when $m_{\tg}\simeq 400$ GeV,
based on 84 pb$^{-1}$ of data taken in Run 1 with $\sqrt{s}=1.8$ TeV.
This analysis assumes eight degenerate squarks when MSSM parameters are used,
and can potentially rule out the case presented here with two species
of nearly degenerate squarks. Thus, we examine the relative signal
produced in the Yukawa unified case with two light squarks with mass
$\sim 90-120$ GeV.
A similar analysis by the D0 collaboration\cite{dzero} 
also excludes $m_{\tq}<250$
GeV (with 10 degenerate squarks) 
by searching for jets$+\eslt$ events, using 79 pb$^{-1}$ of
Run 1 data. This analysis assumes the mSUGRA model, however, which doesn't 
give rise to our case where $m_{\tq}\ll m_{\tg}$.

We generate all sparticle pair production reactions using the event generator
Isajet 7.69, using the model parameters as given in Table~\ref{tab:1},
except allowing $m_{16}(1)$ to float as a free parameter. The mass
value $m_{\tu_R}$ varies nearly linearly with $m_{16}(1)$, so this allows
us to vary the squark masses, while keeping other parameters fixed.
We use the ISAJET
toy detector CALSIM with calorimetry covering the regions $-4<\eta <4$
with cell size $\Delta\eta\times\Delta\phi = 0.1\times
0.262$. Electromagnetic energy resolution is given by $\Delta
E_{em}/E_{em}=0.15/\sqrt{E_{em}}$, while hadronic resolution
is given by $\Delta E_h/E_h=0.7/\sqrt{E_h}$.
Jets are
identified using the ISAJET jet finding algorithm GETJET using a fixed
cone size of $\Delta R=\sqrt{\Delta\eta^2+\Delta\phi^2}=0.7$.
Clusters with
$E_T>15$~GeV and $|\eta (jet)|<2.5$ are labeled as jets.  Muons and
electrons are classified as isolated if they have $E_T>5$~GeV, $|\eta_\ell
|<2.5$, and the visible activity within a cone of $R=0.4$ about the
lepton direction is less than 2 GeV.

\FIGURE{\epsfig{file=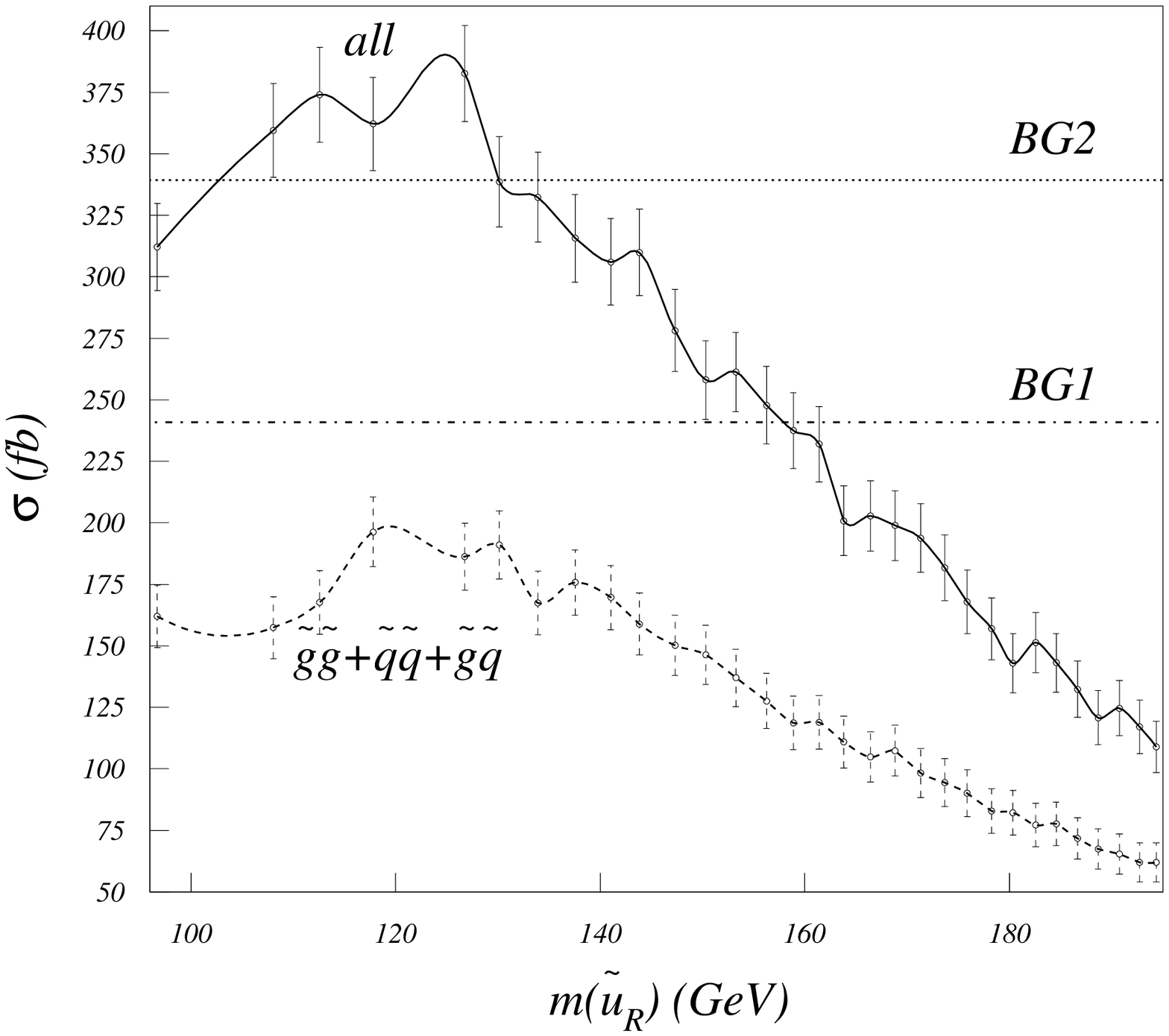,width=14cm} 
\vspace*{-0.8cm}
\caption{Plot of cross section for $\ge 3$ jet$+\eslt$ events $vs.$
$m_{\tu_R}$ at the Fermilab Tevatron collider after implementing CDF cuts,
for same case as in Table~\ref{tab:1}, but with varying $m_{16}(1)$. 
We show plots 
assuming all sparticles contributing to the cross section, and also
the cross section coming just from gluino and squark production. We also show
the 95\% CL signal level for errors added linearly (upper) 
or in quadrature (lower).
}
\label{fig:cdf}
}

The CDF analysis requires the following cuts:
\begin{itemize}
\item $N(jets)\ge 3$,
\item at least one jet has $|\eta |<1.1$,
\item no isolated leptons,
\item the second and third highest $\eslt$ jets are fiducial,
\item $R_1\equiv \sqrt{\delta \phi^2_2+(\pi-\delta \phi^2_1)}>0.75$ rad, 
$R_2\equiv \sqrt{\delta \phi^2_1+(\pi-\delta \phi^2_2)}>0.5$ rad 
($\delta \phi_1=|\phi_{\rm leading\ jet}\it-\phi_{\eslt}|$ 
and $\delta \phi_2=|\phi_{\rm second\ jet}\it-\phi_{\eslt}|$),
\item $\eslt >70$ GeV,
\item $H_T\equiv \eslt +E_T(jet\ 2)+E_T(jet\ 3) \ge150$ GeV,
\end{itemize}
which we also implement into our analysis. In Fig.~\ref{fig:cdf}, 
we plot the signal cross section after cuts.
The dashed curve is the result including
just $\tq\tq +\tg\tq +\tg\tg$ production (as CDF does), while 
the solid curve includes {\it all} sources of $\ge 3$ jet events, including
especially $\tw_1^+\tw_1^-$ and $\tw_1^\pm\tz_2$ production. 
The points and error bars
correspond to our actual signal rate computation, along with statistical error.
We also show the 95\% CL limits as computed by adding maximally allowed
signal cross section with SM error either linearly (BG2) or in quadrature
(BG1). We see that, depending on how the signal and background 
levels are interpreted, squark masses in the 100 GeV range may or may not
be allowed by Run 1 analyses. 
For the case shown, $m_{\tu_R}\sim 100-120$ GeV would be excluded. 
If instead we take a case with larger $m_{1/2}=99.5$ GeV, 
$m_{16}(1)=1297.1$ GeV, $m_{16}(3)=8767.2$ GeV, $m_{10}/m_{16}(3)=1.238$, 
$M_D/m_{16}(3)=0.356$, $\tan\beta =50.79$ and $A_0/m_{16}(3)=-2.13$, 
then we generate a model with $R=1.02$, 
$\Omega_{\tz_1}h^2=0.13$, $m_{\tu_R}=117$ GeV but with
heavier charginos and neutralinos, with $\sigma (3-{\rm jets}+\eslt )=203$ fb,
which is below both BG1 and BG2 levels.
In Fig.~\ref{fig:dots3}, we show model points in the
$m_{\tz_1}\ vs.\ m_{\tu_R}$ plane. In this case, we see the mass gap
$m_{\tu_R}-m_{\tz_1}$ ranging from 25-50 GeV. 
(The bands are due to sampling of $m_{1/2}$ in steps of 5 GeV, while
other parameters are varied continuously.)
Solutions with a large mass gap
are more likely to give large observable rates at Tevatron searches, while 
solutions with small mass gaps will be more difficult to see, since there is
less visible energy released in the squark decays.
\FIGURE{\epsfig{file=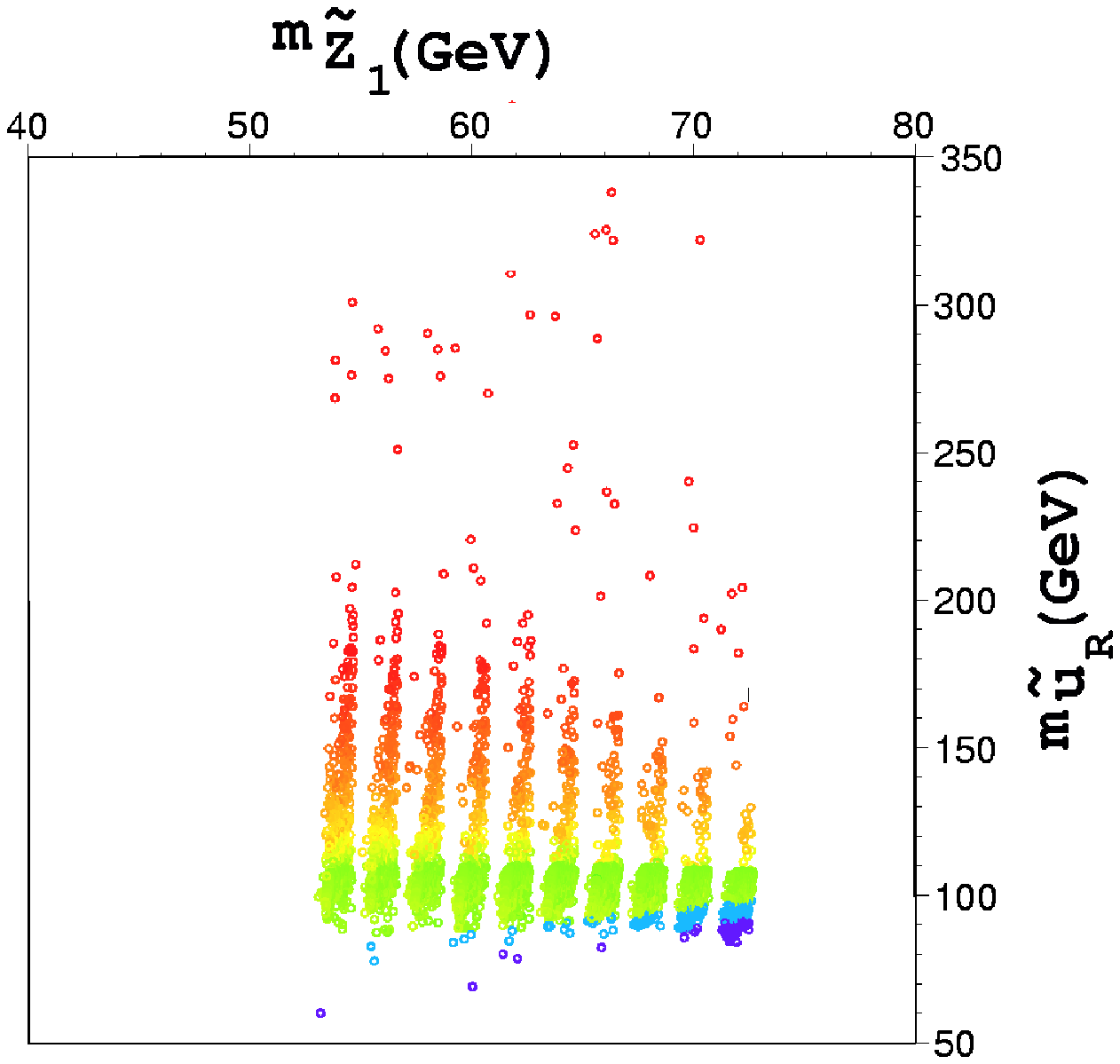,width=10cm}
\vspace*{0.5cm}
\caption{Scan of $SO(10)$ $HS$ models with generational non-universality,
in the $m_{\tu_R}\ vs.\ m_{\tz_1}$ plane. The colors of the dots
correspond to the $\chi^2$ value as listed in Fig.~\ref{fig:dots1}.
}
\label{fig:dots3}
}

An overall assessment seems to be that the two light squark scenario
from Yukawa unified models with generational non-universality is at the edge of
exclusion from previous Tevatron searches using Run 1 data. 
However, given that Run 2 has already
amassed $\sim 400$ pb$^{-1}$ of data, a new dedicated analysis by CDF and/or D0
of $\ge 2-$jets $+\eslt$ events arising from 
squark pair production will likely either rule in or rule out this scenario
with two light squarks with mass in the 90-120 GeV range.

We mention here that there also exist Tevatron searches for
$p\bar{p}\to \tst_1\bar{\tst}_1 X$ production\cite{stop}, where it is
assumed that $\tst_1\to c\tz_1$ 100\% of the time. These analyses
by CDF\cite{cdf_stop} and D0\cite{d0_stop} require a tagged charm jet, 
so would not apply to $\tu_R\bar{\tu}_R$ production. They do apply
to $\tc_R\bar{tc}_R$ production. However, the best limit (by CDF) requires
$m_{\tz_1}< 50$ GeV, which is too light to apply to the scenario
given here: see Fig. \ref{fig:dots3}. 
However, a new search for $c\bar{c} +\eslt$ events 
by CDF and D0 using Run 2 data should be able to probe the parameter
space of the scenario listed in this paper. 

Another consequence of light squarks is that they can mediate
neutralino-nucleon scattering at observable rates via squark exchange in
$s$ and $u$-channel quark-neutralino subprocess diagrams.
In Fig.~\ref{fig:dots4}, we plot various $SO(10)$ $HS$ models
with Yukawa unification to better than 20\%, in the
$\log_{10}\sigma (\tz_1 p)\ vs.\ \Omega_{\tz_1}h^2$ plane. 
We use the neutralino-nucleon scattering rate code developed in
Ref. \cite{ddet}.
The colors
of the dots correspond to the $\chi^2$ value as presented in 
Fig.~\ref{fig:dots1}. We see that the models with low relic density falling
within the WMAP acceptable range also have rather large values of
neutralino-nucleus scattering cross section. In fact, the purple (blue)
dots are already excluded by Edelweiss\cite{edelweiss} 
(CDMS II\cite{cdms2}) searches. 
CDMS II is expected to ultimately probe neutralino-proton scalar
cross sections as low as $10^{-8}$ pb. This essentially covers
the complete range of models with acceptable $\Omega_{\tz_1}h^2$
as obtained from Yukawa unified models with light $\tu_R$ and $\tc_R$
squarks.
\FIGURE{\epsfig{file=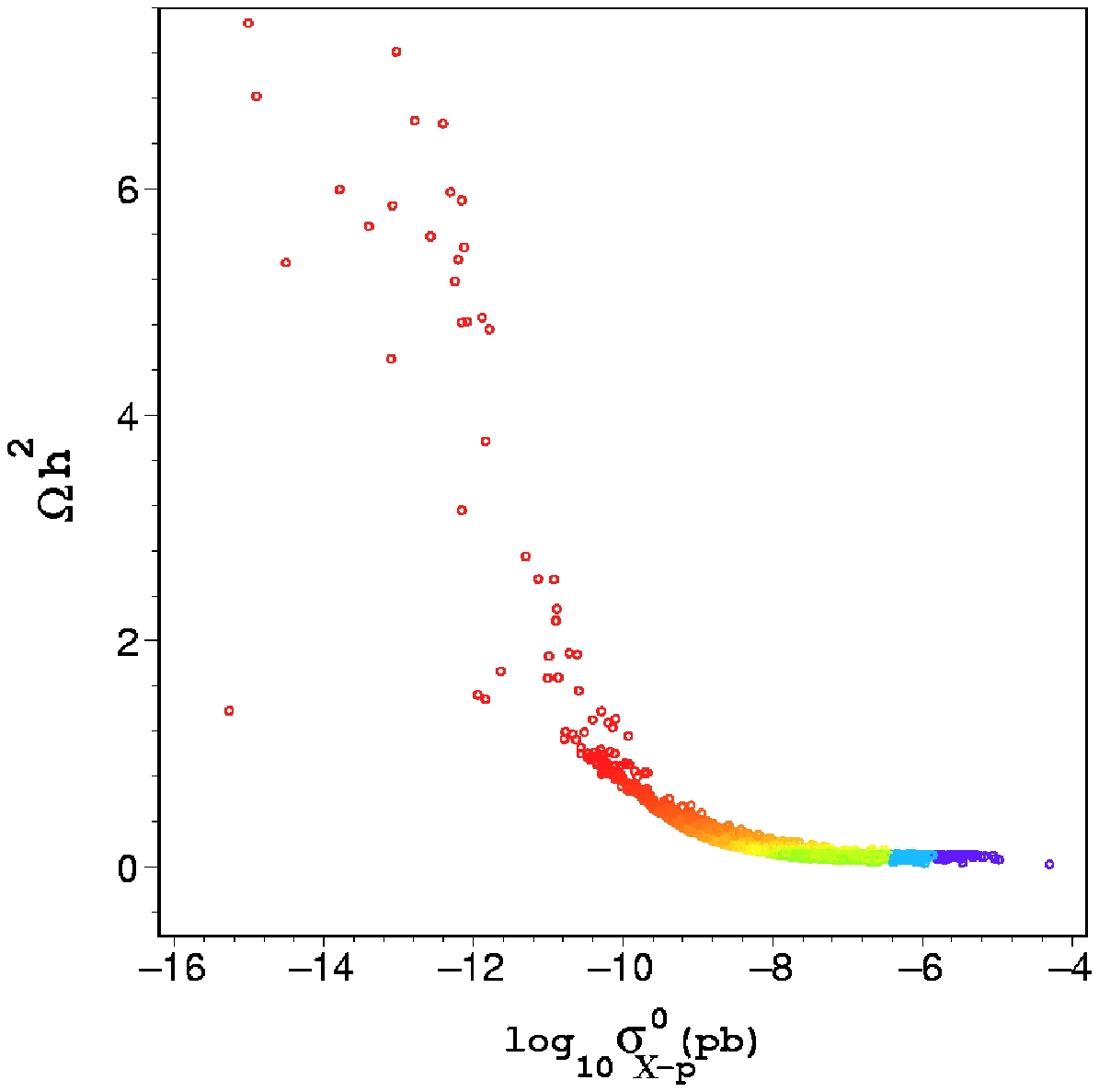,width=10cm}
\vspace*{0.5cm}
\caption{Scan of $SO(10)$ $HS$ models with generational non-universality,
in the $\sigma (\tz_1 p)\ vs.\ \Omega_{\tz_1}h^2$ plane. The colors of the dots
correspond to the $\chi^2$ value computed from
$\Omega_{\tz_1}h^2$, $BF(b\to s\gamma )$ and $a_\mu^{SUSY}$. 
}
\label{fig:dots4}
}

\section{Gaugino mass non-universality}
\label{sec:3}

Aside from generational non-universality in $SO(10)$ models, it is also
possible to have non-universality of gaugino masses.
In a general non-renormalizable SUSY gauge theory, the gaugino masses
arise from the term
\be
{\cal L}\ni -{1\over 4}\int d^2\theta_L f_{AB}(\hat{S})
\overline{\hat{W}_A^c}\hat{W}_B ,
\ee
where we have introduced the gauge kinetic function (GKF) $f_{AB}(\hat{S})$
which is an analytic function of chiral superfields $\hat{S}_i$, and
where $\hat{W}_A$ are gauge superfields. 
In supergravity theories, the GKF can be expanded as
\be
f_{AB}=\delta_{AB}+\frac{\hat{\phi}_{AB}}{M_P}+\cdots ,
\ee
where, to maintain gauge invariance,  
$\hat{\phi}_{AB}$ transforms as the symmetric product of two adjoints.
If the auxiliary field $F_{\phi AB}$ of $\hat{\phi}_{AB}$ obtains
a SUSY breaking (and possibly gauge symmetry breaking) vev, then
gaugino mass terms are induced:
\be
{\cal L}\ni \frac{\langle F_{\phi AB}\rangle}{M_P} 
\overline{\lambda}_A\lambda_B +\cdots  .
\ee
In the case of $SU(5)$ GUT theories\cite{anderson}, the
product of two adjoints can be decomposed as
\be
({\bf 24}\times {\bf 24})_{sym}={\bf 1}+{\bf 24}+{\bf 75}+{\bf 200} ,
\ee
where only the singlet represention leads to universal gaugino masses.
The higher representations lead to relations amongst the
GUT scale gaugino masses which are dictated by group theory, and are 
listed in Ref. \cite{anderson}. In the general case, the field
$\hat{\phi}_{AB}$ could transform as a linear combination of the
above representations, in which case the three MSSM gaugino masses
at $Q=M_{GUT}$ would be essentially free parameters.
In the case of $SO(10)$ GUTS\cite{chamoun,nath}, we have
\be
({\bf 45}\times {\bf 45})_{sym}={\bf 1}+{\bf 54}+{\bf 210}+{\bf 770} .
\ee
\FIGURE{\epsfig{file=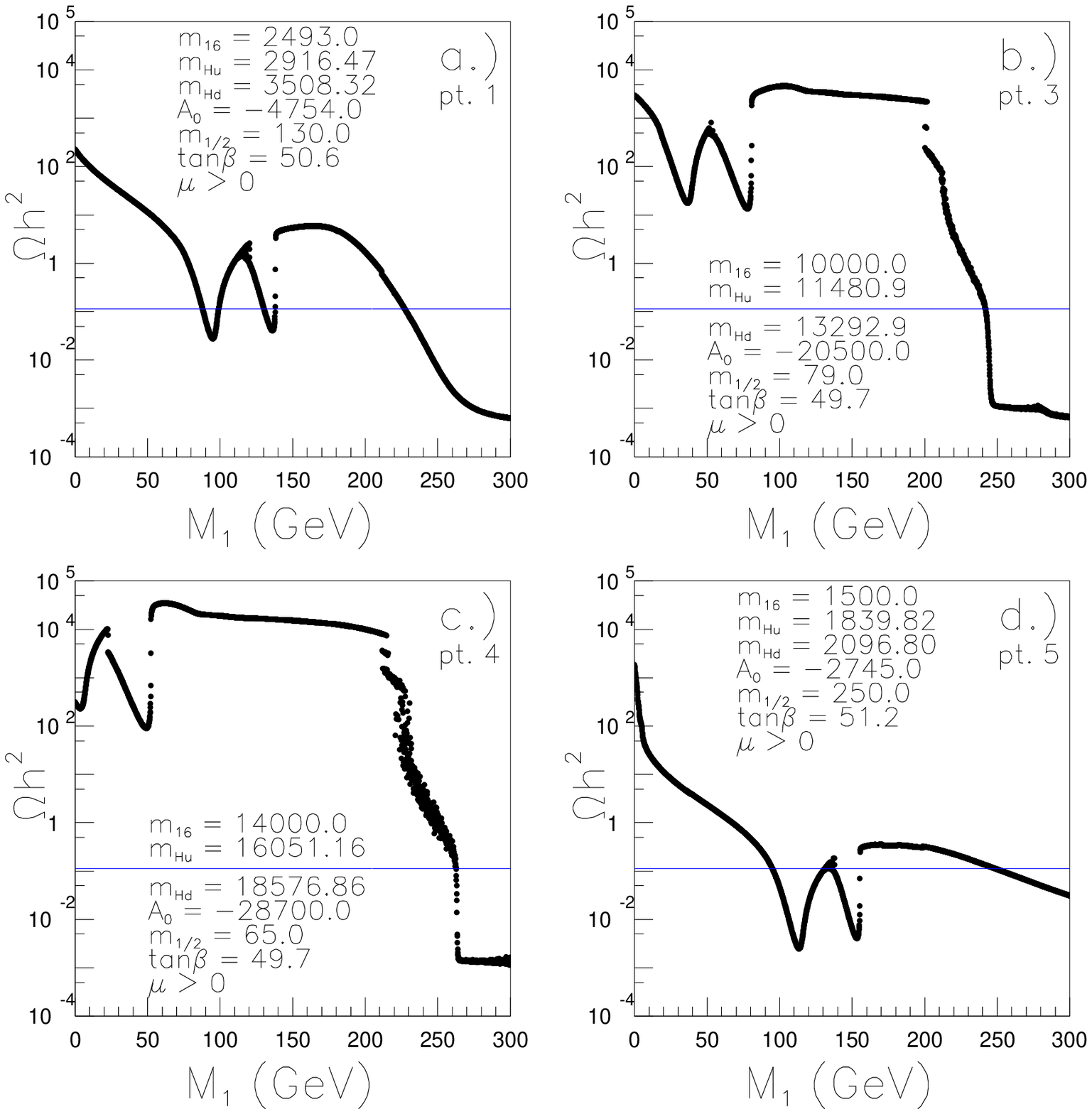,width=14cm} 
\vspace*{-0.8cm}
\caption{Plot of $\Omega_{\tz_1}h^2$ versus GUT scale gaugino mass
$M_1$ for four case studies extracted from Table 1 of 
Ref.~\cite{abbbft}.
We also show as a horizontal line the relic density limit
demanded by WMAP analysis.
}
\label{fig:M_5}
}

For our purposes here, we will not adopt any specific GUT symmetry 
breaking pattern or set of gaugino masses, but instead merely 
enlarge the parameter space by allowing each MSSM gaugino mass as a 
free parameter. Given this freedom, it is then possible to take
any of the Yukawa unified solutions generated by Auto {\it et al.}\cite{abbbft}
and vary the gaugino masses to see if an acceptable relic density solution is 
found. The only problem of concern is whether doing so will destroy the 
Yukawa coupling unification, or lead to models which are violated by 
constraints such as those imposed by LEP2 searches for sparticles.
We note here that an alternative approach is to allow restricted, or 
even complete freedom of gaugino masses, and to explore whether Yukawa 
coupling unification is possible. That approach has been followed in 
the context of 5-d gaugino-mediated SUSY breaking models\cite{5d},
where all scalar masses are set to zero at $M_c\sim M_{GUT}$; 
we refer to the work of Balazs and Dermisek for more details\cite{bd}.

We have examined the positive $\mu$ Yukawa unified solutions
presented in Table 1 of~\cite{abbbft} with regard to variation of
the GUT scale gaugino masses $M_1$ and $M_2$. In the case of
decreasing $M_2$, ultimately the lightest neutralino makes a transition
from being bino-like to being wino-like. Wino-like LSPs have a large
annihilation and co-annihilation rate into SM particles, 
and give a relic density 
almost always much lower than the WMAP bound\cite{birkedal}. 
However, in lowering $M_2$ to
the point of achieving a valid $\Omega_{\tz_1}h^2$,
we find in these five cases that the chargino mass always falls below the LEP2
limits, so that the model becomes excluded by LEP2 rather than WMAP.
The other case, that of varying the $U(1)_Y$ gaugino mass $M_1$, leads to 
models that can satisfy the WMAP bound while being consistent with 
other constraints including those from LEP2.

In Fig.~\ref{fig:M_5}, we show the relic density for four of the points from
Table~1 of~\cite{abbbft} versus variation in GUT scale gaugino mass $M_1$.
In all cases there is a similar structure: for generic values of $M_1$, 
the relic density lies beyond the WMAP limits. However, there exists a 
double dip structure where the relic density may or may not fall below
the WMAP limit, and then at high $M_1$, the relic density drops to very 
low values which would imply non-neutralino cold dark matter.
The high $M_1$ behavior occurs because as $M_1$ is increased, ultimately 
$M_1>M_2$, and the LSP becomes wino-like, although in this case, since $M_2$ 
is fixed, the chargino mass stays fixed beyond the LEP2 mass limits.
At lower $M_1$ values, the double dip structure occurs when the 
value of $2 m_{\tz_1}$ 
drops below the $h$ or the $Z$ resonance. In these cases,
the neutralinos can efficiently annihilate at high rates through these 
resonances in the early universe, leading to lower relic density values.
In some of the cases, the diminution is sufficient to drop
below WMAP limits, while in others (those with very heavy scalars,
and hence with a large $\mu$ parameter),
the diminution is insufficient to save the model.
In each case, we have found that the Yukawa coupling unification
changes by less than a per cent for variation of the $M_1$ parameter
over the range indicated in Fig.~\ref{fig:M_5}.

\section{Conclusions}
\label{sec:4}

Supersymmetric GUTs based on the $SO(10)$ gauge group are highy motivated
theories for physics beyond the standard model. Many of these models
also predict unification of the $t-b-\tau$ Yukawa couplings as well.
The restriction of third generation Yukawa coupling unification 
places stringent constraints on the parameter space of
supersymmetric models, and leads to some new and interesting predictions for
new physics. In this paper, we restrict ourselves to models with
a positive superpotential $\mu$ parameter, which is favored by
$BF(b\to s\gamma )$ and $(g-2)_\mu$ measurements. In these models, 
Yukawa unified sparticle mass solutions with $R<1.05$ can be found, 
but only in the $HS$ case, and for very large $m_{16}\sim 5-15$ TeV
case. This also leads generally to TeV scale values of $m_A$ and $\mu $,
and these numbers, along with TeV scale masses for matter scalars, 
act to suppress
neutralino annihilation in the early universe, and hence to predictions
of a value of $\Omega_{\tz_1}h^2$ far in excess of WMAP 
allowed values.

In this paper, we have explored two methods in which to reconcile
Yukawa unified sparticle mass solutions with the neutralino relic density.
The first case is to allow splitting of the third generation of scalars
from the first two generations. By decreasing the first and second generation
scalar masses, we obtain sparticle mass solutions with very light
$\tu_R$ and $\tc_R$ squark masses, in the 90-120 GeV range. The origin of 
the light squarks comes from the $S$ term in the one-loop RGEs, which 
is non-zero and large in the case of multi-TeV valued split Higgs masses.
A search for just two light squarks at the Fermilab Tevatron collider
using the new Run 2 data should be able to either verify or disprove this
scenario. In addition, the light squarks give rise to large rates for
direct detection of dark matter, and should give observable rates
for searches at CDMS II.

Some possible criticisms of the model presented in Sec. \ref{sec:2} 
is that 1. a degree of fine-tuning is 
required in order to obtain the correct relic density, and 2. the large
trilinear soft breaking $A$ parameters may lead to charge and/or color breaking
minima in the scalar potential. We would concur with the first of these
criticisms. However, our goal in this paper was to examine whether or not 
a relic density in accord with WMAP could be generated in Yukawa unified 
models with $\mu >0$, irregardless of fine-tuning arguments.
Regarding point 2, we note that beginning with Frere {\it et al.}\cite{ccb},
bounds have been derived on soft breaking parameters of the form
$|A_u|^2\le 3(m_Q^2+m_U^2+m_{H_u}^2)$ by trying to avoid
charge and/or color breaking minima of the scalar potential. A complete
analysis must go beyond the tree level and include at least 1-loop
radiative corrections to the scalar potential\cite{casas}. We do not 
include such an analysis here in part because of the philosophical question
of whether false vacua are allowed, and also because a simple solution
of the problem may arise by taking the first two generation $A_0$ parameters
to be light, and non-degenerate with the third generation $A$ parameter.

The other scenario we examined was the case of non-degenerate gaugino masses
at the $GUT$ scale. Beginning with any of the Yukawa unified
solutions found in an earlier study, we find that by dialing $M_1$ to
large enough values, a (partially) wino-like LSP can be generated, with a 
relic density in accord with WMAP allowed values. 
Also, we showed that $M_1$ can
be dialed so that the value of $2m_{\tz_1}$ sits near the $Z$ or $h$
resonance, thus enhancing annihilation rates of the $\tz_1$ in the 
early universe. Depending on the model considered, this may or may not be 
enough to bring the neutralino relic density into accord with the 
WMAP measured value.

\acknowledgments
 
We thank X. Tata for discussions.
This research was supported in part by the U.S. Department of Energy
under contract number DE-FG02-97ER41022.
	
%

\end{document}